\documentclass[aps,twocolumn,prl]{revtex4-1}
\usepackage{graphicx} 
\usepackage{color} 
\usepackage{subfigure}
\usepackage{amsmath} 
\usepackage{amssymb}
\usepackage{bm}
\usepackage{exscale}
\usepackage[mathscr]{eucal}
\usepackage{upgreek}
\usepackage{color}
\usepackage{soul}

\begin{document}

\title{Diffusivity and Hydrodynamic Drag of Nanoparticles at a 
Vapor-liquid Interface}

\author{Joel Koplik$^1$}
\email{jkoplik@ccny.cuny.edu}
\author{Charles Maldarelli$^2$}
\email{cmaldarelli@ccny.cuny.edu}
\affiliation{Benjamin Levich Institute and Departments of Chemical
Engineering$^2$ and Physics$^1$ \\
City College of the City University of New York, New York, NY 10031}

\date{\today}

\begin{abstract} 
Measurements of the surface diffusivity of colloidal spheres translating 
along a vapor/liquid  interface show an unexpected decrease in diffusivity, or increase in surface drag (from the Stokes-Einstein relation) when the 
particles  situate further into the vapor phase. However, 
direct measurements of the surface drag from the colloid  velocity  due to an external force find the expected 
decrease with deeper immersion  into the vapor. The paradoxical drag increase observed in diffusion experiments has been attributed to the attachment of the fluid interface to heterogeneities on the colloid surface, which causes the interface, in response to thermal fluctuations, to either jump or remain pinned, creating added drag. We have performed  molecular dynamics simulations  of the diffusivity and force  experiments for  a nanoparticle with a rough surface at a vapor/liquid interface to examine the effect of contact line fluctuations. The drag calculated from both experiments agree and decrease as the particle positions further into the vapor.  The surface drag is smaller than the bulk liquid drag due to the partial submersion into the liquid, and the finite thickness of the interfacial zone relative to the nanoparticle size. Contact line fluctuations  do not give rise to an anomalous increase in drag.  
\end{abstract} 

\maketitle 

\begin{figure*} 
\centering
\includegraphics[width=.95\textwidth]{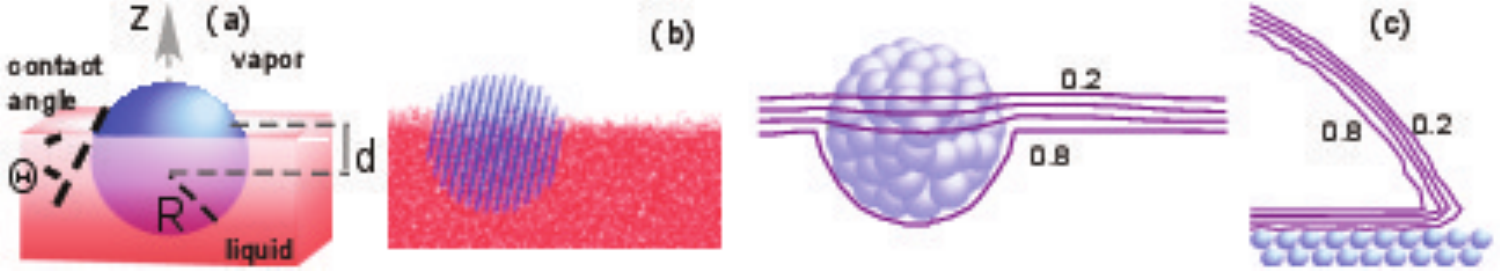}
\caption{\small{(a) Colloid at a vapor/liquid interface, (b) snapshot of the simulated vapor/tetramer interface with the colloid particle and corresponding contour plot for $c=0.9$ and (c) simulation of cylindrical sessile drop of tetramer on solid substrate at $c=0.9$.  In (c) the lines are density contours spaced by 0.2, in units of $\sigma^{-3}$.}}
 \label{Figure1}
\end{figure*}

When a colloid particle breaches the interfacial zone between two 
adjoining immiscible fluid phases the interfacial energy of the particle
changes, because the area of the fluid interface decreases and the contact 
areas of the particle with the bounding fluid phases are altered. For a
particle which only partially wets both adjoining fluids, the change
in interfacial energy is at a minimum when the colloid partially
straddles both adjoining phases. For example, for a spherical colloid
of radius $R$ at a vapor/liquid interface of interfacial tension
$\gamma$ (Fig. \ref{Figure1}a), the  minimum free energy relative to the vapor phase is
\cite{Binks06} $ \Delta \mathcal{F} = - \pi \gamma  R^2 (1 + \cos \theta)^2$ 
where the contact angle  $\theta$ is measured
through the liquid. For particles of large enough size, $\Delta \mathcal{F}$
can overwhelm the typical thermal energy $k_BT$ and
the particle becomes trapped, as thermal fluctuations
cannot dislodge the particle from the interface. Monolayers of
strongly adsorbed colloids at the fluid interfaces of foams and
emulsions provide steric barriers  to coalescence of
the dispersed phase, and find applications as foam and emulsion
stabilizers (e.g. Pickering emulsions). Particle stabilized foams
and emulsions are also used for the fabrication of 
colloid-based solid foams, gels, and bijels, and crystalline
monolayers find applications as superhydrophobic or antireflection
coatings, and templates for micro and nanostructured materials \cite{velev2007}. Central to these
applications is the surface organization of the colloids,
which is a balance between inter-particle attractive and repulsive
forces, external forces applied parallel
to the interface, and the viscous resistance or surface drag due to the
hydrodynamic motion of the colloids along the fluid surface. As we explain, the surface drag is not well understood, with experiments
providing anomalous results, and here we use molecular dynamics to compute and provide insight into the drag.\\
\indent The drag coefficient $\xi$ is the ratio of applied force to resulting velocity (for isolated particles) and is a key parameter in colloidal modeling based on the Langevin equation, for particles moving  either in a bulk liquid or at an interface. For both theory and experiment, two methods are used to calculate the drag. When the particle is allowed to fluctuate in position under the action of Brownian forces alone, the drag is related to the diffusivity $D$ from the ensemble average mean square displacement by the Stokes-Einstein relation $\xi=k_BT/D$. Alternatively, if an external force is applied to the particle to yield a steady velocity, $\xi$ is calculated directly as their ratio. Using the latter method, continuum calculations of the surface drag exerted on a spherical, smooth, nonrotating particle in the limit of zero inertia and 
a flat, zero-thickness interface separating two immiscible liquids have been
undertaken\cite{danov2000viscous, pozrikidis2007particle, fischer2006viscous, dorr2016}. The
continuum drag increases  with  immersion into the more
viscous phase, and approaches the Stokes bulk drag coefficient $6\pi\mu R$, where $\mu$ is the fluid viscosity 
far from the surface.  The drag on smaller, nanometer-sized 
colloidal particles has been addressed theoretically,
using molecular dynamics (MD) simulations of spherical rigid or structureless
particles at an atomistically fluctuating liquid interface to calculate the 
surface diffusivity from the mean square displacement. For liquid/vapor
interfaces in Lennard-Jones (LJ) systems \cite{Cheung2010, grest2012, Shahab2015}, 
the expected increase of surface diffusivity with displacement into the vapor is found, 
and for a water/polydimethylsiloxane interface\cite{dai2009a} the diffusivity was
intermediate between the simulated bulk diffusion coefficients. \\ 
\indent Experiments have directly obtained the surface drag on large spherical
colloids (1 - 10$^{3}$ $\mu$m in diameter) at gas/aqueous and
oil/aqueous interfaces by moving the particles with magnetic\cite{ally2010magnetophoretic} or
capillary forces (e.g.\cite{petkov1995measurement, dalbe2011aggregation}),  and measuring the resultant velocity using optical
microscopy and particle tracking. To compare to
theory, the particle immersion depth is evaluated separately by a
measurement of the contact angle, and the measured surface drag is
in agreement with the continuum predictions.  For spherical colloids approximately one  $\mu$m in diameter at oil/water interfaces, measurements of the surface diffusivity and drag from the Stokes-Einstein equation are in agreement\cite{tong2009, Berglund2012} with continuum theory. But for these same particles 
at an air/water interface or  nanometer sized particles at an oil/water interface, 
the surface diffusivity is smaller than would have been predicted from the Stokes
Einstein equation using the continuum surface drag, or, paradoxically decreases with crossing into the gas or less viscous phase, becoming smaller than the bulk diffusivity\cite{Stone2007, tong2008, koynov2011, fisher2014,boniello2015}.\\
\indent Several mechanisms have been proposed to explain the unexpectedly low 
surface diffusivities.  Obviously, surfactant 
contaminants could give rise to viscous surface shear and  Marangoni forces which increase
the surface drag. However, in diffusion experiments in which tension measurements
indicate  a relatively clean interface, studies suggest that  thermal fluctuations at the contact line can create forces on the particle\cite{boniello2015}. In particular, the surface of colloids are typically not smooth, and the fluid interface can become pinned at heterogeneities causing hysteresis in the measurement of the contact angle\cite{kaz2012physical}. As a particle moves along a fluid interface, thermal fluctuations of the interface can cause a contact line pinned to a heterogeneity to hop to an adjacent heterogeneity. Alternatively, if the pinning is strong enough, the contact line stays in place during motion and the interface distorts.  In either case changes in the interface slope creates forces on the particle  which are balanced by additional drag. Molecular dynamics calculations can simulate contact line fluctuations, and our objective is to use MD for an atomistic particle with surface roughness at a vapor/liquid interface to study whether these fluctuations can account for the anomalously large surface drag measured in diffusion experiments.  Since the constant force experiments did not show this anomaly, and to provide self-consistent results, we calculate the drag coefficient for the cases in which the particle is allowed to fluctuate by Brownian forces alone, or the particle is subject to an external force and we undertake these for different immersion depths  or particle wettabilities.\\
\indent The simulations employ basic MD methods.\cite{frenkelsmit}  We consider a liquid bath of tetramer molecules, composed of FENE chains of Lennard-Jones atoms of core diameter $\sigma$, energy scale $\epsilon$ and mass $m$, interacting via $\displaystyle{
V_{\rm LJ}(r) = 4\,\epsilon\, \left[ (r/\sigma)^{-12}
- c\, (r/\sigma)^{-6}\ \right]}$ and define a time scale $\tau=\sigma(m/\epsilon)^{1/2}$. The parameter $c$ adjusts the strength of
the attractive interaction; $c=1$ for tetramer interactions and is varied for the interaction of the liquid atoms with the atoms of the colloid particle to adjust the wettability (the contact angle) and thereby vary the immersion depth. The simulation cell is in the form of a slab with a free surface at the top and in contact with a bottom consisting of two layers of the same LJ atoms attached by linear tether springs to fcc lattice sites. The bottom prevents free translation of the system and periodic boundary conditions are imposed in the lateral directions. A local Nose-Hover 
thermostat fixes the temperature at $0.8\epsilon/k_{B}$, for which the tetramer liquid and vapor are in equilibrium with a vapor/liquid surface tension (obtained by numerical integration of the the difference between normal and transverse stress across the interface) $\gamma=0.668\epsilon/\sigma^2$. The bulk liquid density is 0.857$\sigma^{-3}$ and viscosity is $\eta=5.18m/(\sigma \tau)$, obtained from a separate simulation of the same liquid placed between two solid walls in Couette flow.  The solid particle is a rigid spherical section of a cubic lattice of LJ atoms at the same density as the equilibrated liquid, formed by enclosing all atoms of the lattice within a radius
$8\sigma$ of a central atom, which yields a distinctly rough surface. Further details are in the Supplementary Information\cite{suppl}. \\ 
\indent We first compute the immersion depth $d$ (Fig. \ref{Figure1}a) from simulations of diffusion. The particle is initially placed below the interface and allowed to migrate upwards to its equilibrium position and diffuse there. The particles are not explicitly constrained to lie at the interface, and move according to Newton's and Euler's equations and do exhibit some vertical motion, but the typical height fluctuation in small with a standard deviation in the height distribution of at most $0.6 \sigma$, even for the lowest wettability. The immersion depth is based on the average horizontal position of the center of the particle relative to planes of iso-contours of the fluid density far from the interface, see Fig.~\ref{Figure1}b for $c=0.9$. The contours make clear that the particle diameter $R$ is comparable to the size of the interfacial transition zone.  When the particle radius is much larger than the zone thickness, contours of different density intersect the particle at the same angle, defining a unique contact angle and immersion depth. Here, the contours intersect at different angles and the contact angle is ambiguous. We choose to define the immersion depth as the vertical distance from  
\begin{figure}  
\includegraphics[width=0.50\textwidth]{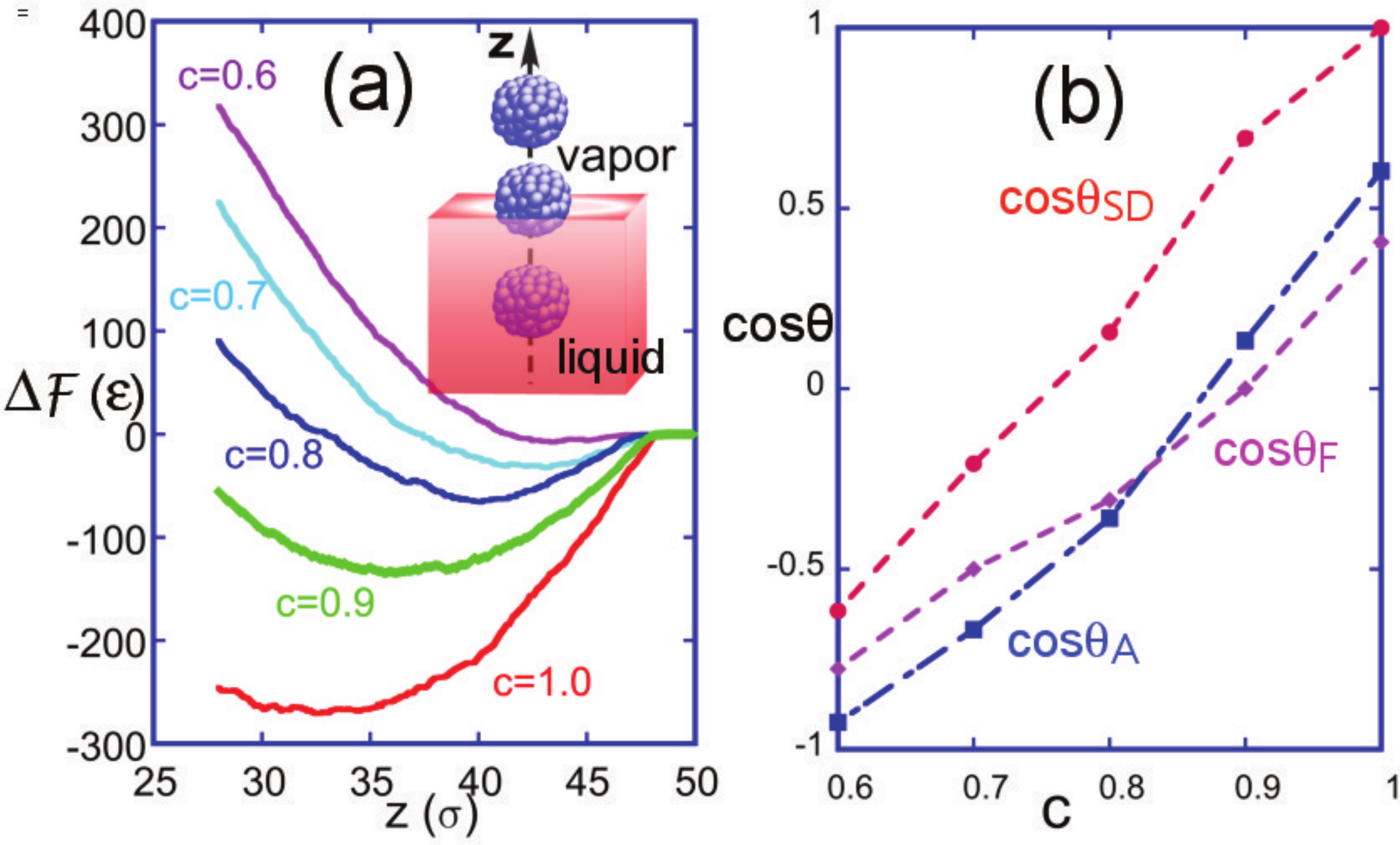} 
\caption{\small{(a) Immersion free energy calculation and (b) contact angle (cos$\theta$) as a function of c obtained from simulation ($\theta_{A}$), from a cylindrical sessile drop simulation ($\theta_{SD}$) and from free energy calculation ($\theta_{F}$).}}
\label{Figure2}
\end{figure}
the half-bulk-density contour to the particle center, defining an apparent contact angle ($d/R=\rm{cos}\theta_{A}$). The immersion depth and apparent angle are  plotted in Fig. \ref{Figure2}b as a function of $c$.  $\theta_{A}$ can be compared to the angle measured directly in analogous MD simulations of sessile drops of the tetramer on a planar substrate of atoms identical to the colloid and in the same lattice configuration (Fig. \ref{Figure1}c for c=0.9). The large substrate area allows the fluid interface contours to intersect it as a set of concentric circles,  at nearly identical angles $\theta_{SD}$, plotted alongside the apparent angles in Fig. \ref{Figure2}b. Note $\theta_{SD}<\theta_{A}$, and can only be brought into congruence by defining $d$ relative to a contour of lower density. \\
\begin{figure*}  
\includegraphics[width=0.99\textwidth]{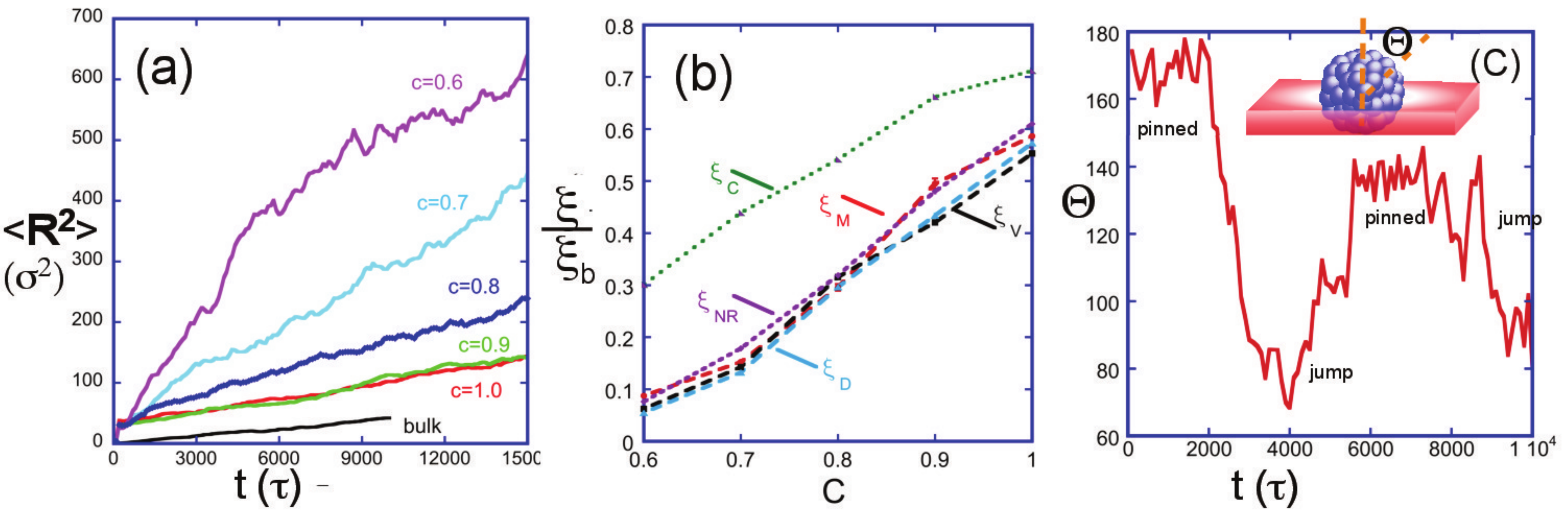}
 \caption{(a) Mean-square displacement in the plane of the
  interface {\em vs}. time, for various wettabilities. (b) surface drag relative to bulk drag  $\xi/\xi_{b}$ as a function of interaction parameter $c$, and (c) sample rotation angle $\theta$ during a diffusion simulation (c=0.7).}
  \label{Figure3}
 \end{figure*}
\indent A second measure of the equilibrium immersion depth can be obtained by calculating the free energy of the system as a function of the vertical position of the particle by thermodynamic integration\cite{sepideh2013}. Since temperature is held fixed, the change in Helmholtz free energy  between two immersion depths equals the work done when the particle is moved quasi-statically and reversibly between the two points. Explicitly, $\Delta \mathcal{F}= \int \bf{F(r)}\cdot d\bf{r}$, where $\bf{F}$ is the force on the particle at $\bf{r}$.  
To evaluate the integral, the liquid slab is equilibrated until the
density field stabilizes. The particle is first placed in the nearly-empty region above the interface and 
slowly displaced downward into the liquid, alternately moving it by  $0.1\sigma$ at a velocity 0.001 $\sigma/\tau$, re-equilibrating the system for 50$\tau$ and then averaging the force while it is fixed in position for $100\tau$.  The results are given in 
Fig.~\ref{Figure2}a for the various wettabilities.  In each case, including the completely wetting value
$c=1$, $\Delta \mathcal{F}$ is zero until the particle contacts the fluid, then dips reflecting the attraction of the liquid atoms,
and then displays a minimum inside the fluid at a position that deepens as the wettability increases. The curves flatten out
at lower values of $z$ fully inside the bulk liquid. The minimum is quite shallow for the case $c=0.6$ where the particle is barely in contact with the fluid, indicating weak binding. A contact angle and immersion depth based on the free energy minimum, $\Delta \mathcal{F}/(\pi R^2\gamma)= -(1+\rm{cos}\theta_{F})^2$, is also plotted in Fig. \ref{Figure2}b; note that the immersion into the liquid increases with the strength of the liquid-solid interaction, $c$  in agreement with MD calculations of Cheng and Grest\cite{grest2012}.\\  
\indent We begin our calculation of the surface drag coefficient by validating our simulations for a completely immersed colloid ($c$=1) in a constant force simulation. The colloid is placed in the center of a slab of liquid confined between two walls identical to the bottom wall of our vapor/liquid simulation cell and with the spacing adjusted to give the 
same bulk liquid density as when a liquid-vapor interface is present. We translate the colloid at a 
a constant velocity through the middle of the channel and measuring the net force the fluid exerts on it.  
More precisely, at each time step $\Delta t$ the center of mass of the particle is translated by ${\bm U}\Delta t$ while the particle is allowed to 
rotate according to the Euler equations in response to the net torque exerted  on it by fluid atoms. The net force $\bm{F}_{ext}$ exerted by the fluid on the particle is recorded (but not used to update the center of mass position)  and averaged over the duration of the simulation.  Although the force  fluctuates strongly from time step to time step, its average 
is quite stable when averaged over a time interval of 100$\tau$ or more.  We define the drag coefficient $\xi$
via ${\bm{F}}_{\rm{ext}} = \xi {\bm{U}}$, and a bulk run of 1500$\tau$ yields $\xi_{b}=(756\pm
30) m/\tau$.  The result is in good agreement with the expected drag 
coefficient for a perfect no-slip sphere in low Reynolds number flow in an
unbounded fluid, $\xi_{b}=6\pi\mu R = 781 m/\tau$. Wall effects are negligible for a sphere of
radius 8$\sigma$ at the center of a channel of width 60$\sigma$: the
correction to $\xi_{b}$ is only about 0.2\% \cite{Pasol2011}, and comparable to the 
statistical error.  The discrepancy may be attributed to the particle's surface roughness, which introduces some uncertainty in the definition of
radius;  we could define a  hydrodynamic radius equal to  $7.74 \sigma$.     Likewise, the periodicity in the other two directions
would modify the result but we expect this effect to be similarly small.\\
\indent We next compute $\xi$ from simulations of the diffusivity of a completely immersed colloid using the same simulation cell of a liquid between two walls. In this case, the colloid is allowed to migrate to the interface and move freely according to Newton's law, based on the instantaneous force exerted by the fluid with no external forcing.  We consider a statistical ensemble of 90 independent realizations, with different random
seeds used to generate the initial velocity distribution, and record the average mean square displacement as a function of time for simulations of $1.5\times10^{4}\tau$. In order to have a
close comparison with the cases of particles at an interface where only
diffusion in the plane of the interface is relevant, and to minimize the
effect of the confining walls, we focus on the
two-dimensional diffusivity $D$ in the plane parallel to the channel walls, 
which is related to the mean square displacement by
$\langle {\bf R}^2\rangle \equiv \langle x^2+y^2\rangle = 4Dt$.
The result, the black lower curve in Fig.~\ref{Figure3}a, is a noisy 
straight line whose slope is determined by a least
squares fit to be $D_{b}=0.0012 \sigma^2/\tau$ which when inserted 
into the Stokes-Einstein relation $\xi=k_BT/D$ predicts a drag coefficient 
$\xi_{b}=(788\pm 9)m/\tau$, again in good agreement with the constant external force 
drag calculation and the continuum prediction. We also record the instantaneous 
particle velocity as a function of time and compute the in-plane velocity 
autocorrelation function (VACF) $c(t)=\frac{1}{2}\langle\langle
v_x(t+\tau)\,v_x(\tau)+v_y(t+\tau)\,v_y(\tau)\rangle\rangle$, where the 
brackets
refer to an average over realizations and over the starting time $\tau$.
Mathematically, $\displaystyle{D=\lim_{t\to\infty} \frac{1}{4t}\langle {\bf R}^2(t)\rangle
=\lim_{t\to\infty}\int_0^t dt'\,c(t')}$,
but the two expressions treat the data differently and need not quite agree
for finite samples and finite measurement times.  In practice the integral
of the VACF first increases as a function of the upper limit, then reaches a
plateau whose value is taken to be $D$, and then oscillates at larger values 
of $t$ where the VACF fluctuates about zero. 
The resulting estimate, $D_{b}=0.0011\sigma^2/\tau$ and $\xi_{b}=792$,
is quite consistent with the previous values.   
Lastly, we note that the intercept of the VACF, $c(0)=4.60\times
10^{-4}(\sigma/\tau)^2$, 
agrees with the value $k_BT/M$ ($M$ the colloid mass) expected from the equipartition theorem,
and furthermore the probability distributions for the three particle velocity 
components are Gaussians with this width.\\
\indent We now calculate the surface drag coefficient $\xi$ from diffusion and constant force simulations for colloids at an interface for different immersion depths.  For the diffusion simulation, the two-dimensional mean-square displacement vs. time 
is given in Fig.~\ref{Figure3}a. As in the fully-immersed case, the slope is
determined by a least-squares fit and converted into a drag coefficient
using the Stokes-Einstein relation, and the results are recorded in Fig. \ref{Figure3}b as $\xi_{M}/\xi_{b}$ as a function of $c$.  
We also measure the velocity autocorrelation in these simulations and its time integral provides
a second, semi-independent determination of the diffusivity, and is in good agreement with $\xi_{M}$ (\ref{Figure3}b, $\xi_{V}$). For the the calculation of $\xi$ by the application of a constant force, for each wettability or value $c$, the particle is fixed at the mean height obtained in the diffusion calculation, and dragged parallel to the interface at a
fixed velocity 0.1$\sigma/\tau$ while allowed to rotate freely, and the 
ratio of force to velocity is recorded in Fig. \ref{Figure3}b as $\xi_D$. The diffusion and constant force measures of $\xi$ are within ten percent agreement. The surface drag decreases as the particle displaces into the vapor phase and is always less than the drag in the bulk liquid. Importantly, the anomalously large drag which has been obtained when the particle situates further into the gas phase is not observed.\\
\indent To examine the contact line pinning and depinning during the simulations, which as we have noted has been conjectured as a reason for anomalously large drag, we have monitored the orientation of the particle with time, 
both when the colloid is simply diffusing or being dragged. To do this, we define a director (Fig. \ref{Figure3}c, inset) as the
unit vector between two atoms at opposite sides of the (rigid) particle and 
record its orientation angle as a function of time, both with respect to the 
interface normal ($\theta$) and in the plane of the interface ($\phi$).  Visualizations of the interface as Fig. \ref{Figure1}b show a relatively planar surface intersecting the colloid, so that the body fixed angle $\theta$ locates the contact line. In individual diffusion runs, while $\phi$ tends to vary randomly, $\theta$ (see Fig. \ref{Figure3}c), varies erratically, showing either intervals of continuous variation (no pinning or hopping), and intermittent jumps  to orientations where it remains pinned. While expected for a rough surface, this contact line movement does not elevate the drag. To examine the limiting case of strong pinning in which the particle does not rotate in  $\phi$ or $\theta$, we repeated the constant force simulation but did not permit the particle to rotate. We find the drag ($\xi_{NR}$) to be expectedly larger than in the case of free rotation, but only marginally so (see Fig. \ref{Figure3}b) and this restriction again does not  lead to an anomalously large drag. The Stokes continuum drag of a nonrotating smooth  particle moving along a  gas/liquid interface\cite{danov2000viscous, pozrikidis2007particle, fischer2006viscous, dorr2016}  is shown in Fig. \ref{Figure3}b as $\xi_c$ and is larger than the drag computed from the MD simulations.  The interfacial width is the likely source of the discrepancy: the sphere in these simulations is partly in contact with lower density fluid in the interfacial zone which exerts less force, as has also been found by  \cite{Cheung2010, grest2012}.\\
\indent To conclude,  in all cases of nanoparticles with a rough surface  at a vapor/liquid 
interface, the drag coefficient inferred from diffusion
MD simulations and  the Stokes-Einstein relation are are in agreement with  the drag
coefficient $\xi$ obtained from the constant force MD simulations, and these show that the drag decreases as the colloid becomes more immersed in the gas phase. Since the size of the nanoparticle was comparable to the thickness of  interfacial zone, the surface drags were  less than values computed from continuum studies with a sharp interface.  While contact line pinning and de-pinning is observed in the diffusion simulations, they do not create large surface drag as underscored by the fact that MD simulations for a pinned non-rotating colloid dragged across the surface has a surface drag only slightly larger than the free-rotating case.
\begin{acknowledgments}
JK was supported in part by NSF grant CBET-1264550.
\end{acknowledgments}

\begin{thebibliography}{25}%
\makeatletter
\providecommand \@ifxundefined [1]{%
 \@ifx{#1\undefined}
}%
\providecommand \@ifnum [1]{%
 \ifnum #1\expandafter \@firstoftwo
 \else \expandafter \@secondoftwo
 \fi
}%
\providecommand \@ifx [1]{%
 \ifx #1\expandafter \@firstoftwo
 \else \expandafter \@secondoftwo
 \fi
}%
\providecommand \natexlab [1]{#1}%
\providecommand \enquote  [1]{``#1''}%
\providecommand \bibnamefont  [1]{#1}%
\providecommand \bibfnamefont [1]{#1}%
\providecommand \citenamefont [1]{#1}%
\providecommand \href@noop [0]{\@secondoftwo}%
\providecommand \href [0]{\begingroup \@sanitize@url \@href}%
\providecommand \@href[1]{\@@startlink{#1}\@@href}%
\providecommand \@@href[1]{\endgroup#1\@@endlink}%
\providecommand \@sanitize@url [0]{\catcode `\\12\catcode `\$12\catcode
  `\&12\catcode `\#12\catcode `\^12\catcode `\_12\catcode `\%12\relax}%
\providecommand \@@startlink[1]{}%
\providecommand \@@endlink[0]{}%
\providecommand \url  [0]{\begingroup\@sanitize@url \@url }%
\providecommand \@url [1]{\endgroup\@href {#1}{\urlprefix }}%
\providecommand \urlprefix  [0]{URL }%
\providecommand \Eprint [0]{\href }%
\providecommand \doibase [0]{http://dx.doi.org/}%
\providecommand \selectlanguage [0]{\@gobble}%
\providecommand \bibinfo  [0]{\@secondoftwo}%
\providecommand \bibfield  [0]{\@secondoftwo}%
\providecommand \translation [1]{[#1]}%
\providecommand \BibitemOpen [0]{}%
\providecommand \bibitemStop [0]{}%
\providecommand \bibitemNoStop [0]{.\EOS\space}%
\providecommand \EOS [0]{\spacefactor3000\relax}%
\providecommand \BibitemShut  [1]{\csname bibitem#1\endcsname}%
\let\auto@bib@innerbib\@empty
\bibitem [{\citenamefont {Binks}\ and\ \citenamefont
  {Horozov}(2006)}]{Binks06}%
  \BibitemOpen
  \bibinfo {editor} {\bibfnamefont {B.}~\bibnamefont {Binks}}\ and\ \bibinfo
  {editor} {\bibfnamefont {T.}~\bibnamefont {Horozov}},\ eds.,\ \href@noop {}
  {\emph {\bibinfo {title} {Colloidal Particles at Liquid Interfaces}}}\
  (\bibinfo  {publisher} {Cambridge Univ. Press},\ \bibinfo {address} {Cambridge,
  UK},\ \bibinfo {year} {2006})\BibitemShut {NoStop}%
\bibitem [{\citenamefont {Prevo}\ \emph {et~al.}(2007)\citenamefont {Prevo},
  \citenamefont {Kuncicky},\ and\ \citenamefont {Velev}}]{velev2007}%
  \BibitemOpen
  \bibfield  {author} {\bibinfo {author} {\bibfnamefont {B.}~\bibnamefont
  {Prevo}}, \bibinfo {author} {\bibfnamefont {D.}~\bibnamefont {Kuncicky}}, \
  and\ \bibinfo {author} {\bibfnamefont {O.}~\bibnamefont {Velev}},\
  }\href@noop {} {\bibfield  {journal} {\bibinfo  {journal} {Colloids
  Surfaces A}\ }\textbf {\bibinfo {volume} {311}},\ \bibinfo {pages}  and{2}
  (\bibinfo {year} {2007})}\BibitemShut {NoStop}%
\bibitem [{\citenamefont {Danov}\ \emph {et~al.}(2000)\citenamefont {Danov},
  \citenamefont {Dimova},\ and\ \citenamefont {Pouligny}}]{danov2000viscous}%
  \BibitemOpen
  \bibfield  {author} {\bibinfo {author} {\bibfnamefont {K.~D.}\ \bibnamefont
  {Danov}}, \bibinfo {author} {\bibfnamefont {R.}~\bibnamefont {Dimova}}, \
  and\ \bibinfo {author} {\bibfnamefont {B.}~\bibnamefont {Pouligny}},\
  }\href@noop {} {\bibfield  {journal} {\bibinfo  {journal} {Phys. 
  Fluids}\ }\textbf {\bibinfo {volume} {12}},\ \bibinfo {pages} {2711}
  (\bibinfo {year} {2000})}\BibitemShut {NoStop}%
\bibitem [{\citenamefont {Pozrikidis}(2007)}]{pozrikidis2007particle}%
  \BibitemOpen
  \bibfield  {author} {\bibinfo {author} {\bibfnamefont {C.}~\bibnamefont
  {Pozrikidis}},\ }\href@noop {} {\bibfield  {journal} {\bibinfo  {journal}
  { J. Fluid Mech. }\ }\textbf {\bibinfo {volume} {575}},\ \bibinfo {pages} 	{333} (\bibinfo
  {year} {2007})}\BibitemShut {NoStop}%
\bibitem [{\citenamefont {Fischer}\ \emph {et~al.}(2006)\citenamefont
  {Fischer}, \citenamefont {Dhar},\ and\ \citenamefont
  {Heinig}}]{fischer2006viscous}%
  \BibitemOpen
  \bibfield  {author} {\bibinfo {author} {\bibfnamefont {T.~M.}\ \bibnamefont
  {Fischer}}, \bibinfo {author} {\bibfnamefont {P.}~\bibnamefont {Dhar}}, \
  and\ \bibinfo {author} {\bibfnamefont {P.}~\bibnamefont {Heinig}},\
  }\href@noop {} {\bibfield  {journal} {\bibinfo  {journal} {J. Fluid             	Mech. }\ }\textbf
  {\bibinfo {volume} {558}},\ \bibinfo {pages} {451 } (\bibinfo {year}
  {2006})}\BibitemShut {NoStop}%
\bibitem [{\citenamefont {Dorr}\ \emph {et~al.}(2016)\citenamefont {Dorr},
  \citenamefont {Hardt}, \citenamefont {Masoud},\ and\ \citenamefont
  {Stone}}]{dorr2016}%
  \BibitemOpen
  \bibfield  {author} {\bibinfo {author} {\bibfnamefont {A.}~\bibnamefont
  {Dorr}}, \bibinfo {author} {\bibfnamefont {S.}~\bibnamefont {Hardt}},
  \bibinfo {author} {\bibfnamefont {H.}~\bibnamefont {Masoud}}, \ and\ \bibinfo
  {author} {\bibfnamefont {H.}~\bibnamefont {Stone}},\ }\href@noop {}
  {\bibfield  {journal} {\bibinfo  {journal} {J. Fluid Mech.}\
  }\textbf {\bibinfo {volume} {790}},\ \bibinfo {pages} {607} (\bibinfo {year}
  {2016})}\BibitemShut {NoStop}%
\bibitem [{\citenamefont {Cheung}(2010)}]{Cheung2010}%
  \BibitemOpen
  \bibfield  {author} {\bibinfo {author} {\bibfnamefont {D.}~\bibnamefont
  {Cheung}},\ }\href {\doibase http://dx.doi.org/10.1016/j.cplett.2010.06.074}
  {\bibfield  {journal} {\bibinfo  {journal} {Chem. Phys. Lett.}\
  }\textbf {\bibinfo {volume} {495}},\ \bibinfo {pages} {55 } (\bibinfo {year}
  {2010})}\BibitemShut {NoStop}%
\bibitem [{\citenamefont {Cheng}\ and\ \citenamefont
  {Grest}(2012)}]{grest2012}%
  \BibitemOpen
  \bibfield  {author} {\bibinfo {author} {\bibfnamefont {S.}~\bibnamefont
  {Cheng}}\ and\ \bibinfo {author} {\bibfnamefont {G.}~\bibnamefont {Grest}},\
  }\href@noop {} {\bibfield  {journal} {\bibinfo  {journal} {Journal of Chemical Phyi}\
  }\textbf {\bibinfo {volume} {136}},\ \bibinfo {pages} {214702} (\bibinfo
  {year} {2012})}\BibitemShut {NoStop}%
\bibitem [{\citenamefont {Rezvantalab}\ \emph {et~al.}(2015)\citenamefont
  {Rezvantalab}, \citenamefont {Drazer},\ and\ \citenamefont
  {Shojaei-Zadeh}}]{Shahab2015}%
  \BibitemOpen
  \bibfield  {author} {\bibinfo {author} {\bibfnamefont {H.}~\bibnamefont
  {Rezvantalab}}, \bibinfo {author} {\bibfnamefont {G.}~\bibnamefont {Drazer}},
  \ and\ \bibinfo {author} {\bibfnamefont {S.}~\bibnamefont {Shojaei-Zadeh}},\
  }\href@noop {} {\bibfield  {journal} {\bibinfo  {journal} {J. Chem. Phs.}\
  }\textbf {\bibinfo {volume} {142}},\ \bibinfo {pages} {014701} (\bibinfo
  {year} {2015})}\BibitemShut {NoStop}%
\bibitem [{\citenamefont {Song}\ \emph {et~al.}(2009)\citenamefont {Song},
  \citenamefont {Luo},\ and\ \citenamefont {Dai}}]{dai2009a}%
  \BibitemOpen
  \bibfield  {author} {\bibinfo {author} {\bibfnamefont {Y.}~\bibnamefont
  {Song}}, \bibinfo {author} {\bibfnamefont {M.}~\bibnamefont {Luo}}, \ and\
  \bibinfo {author} {\bibfnamefont {L.}~\bibnamefont {Dai}},\ }\href@noop {}
  {\bibfield  {journal} {\bibinfo  {journal} {Langmuir}\ }\textbf {\bibinfo
  {volume} {26}},\ \bibinfo {pages} {5} (\bibinfo {year} {2009})}\BibitemShut
  {NoStop}%
\bibitem [{\citenamefont {Ally}\ and\ \citenamefont
  {Amirfazli}(2010)}]{ally2010magnetophoretic}%
  \BibitemOpen
  \bibfield  {author} {\bibinfo {author} {\bibfnamefont {J.}~\bibnamefont
  {Ally}}\ and\ \bibinfo {author} {\bibfnamefont {A.}~\bibnamefont
  {Amirfazli}},\ }\href@noop {} {\bibfield  {journal} {\bibinfo  {journal}
  {Colloids and Surfaces A}\ }\textbf
  {\bibinfo {volume} {360}},\ \bibinfo {pages} {120} (\bibinfo {year}
  {2010})}\BibitemShut {NoStop}%
\bibitem [{\citenamefont {Petkov}\ \emph {et~al.}(1995)\citenamefont {Petkov},
  \citenamefont {Denkov}, \citenamefont {Danov}, \citenamefont {Velev},
  \citenamefont {Aust},\ and\ \citenamefont {Durst}}]{petkov1995measurement}%
  \BibitemOpen
  \bibfield  {author} {\bibinfo {author} {\bibfnamefont {J.~T.}\ \bibnamefont
  {Petkov}}, \bibinfo {author} {\bibfnamefont {N.~D.}\ \bibnamefont {Denkov}},
  \bibinfo {author} {\bibfnamefont {K.~D.}\ \bibnamefont {Danov}}, \bibinfo
  {author} {\bibfnamefont {O.~D.}\ \bibnamefont {Velev}}, \bibinfo {author}
  {\bibfnamefont {R.}~\bibnamefont {Aust}}, \ and\ \bibinfo {author}
  {\bibfnamefont {F.}~\bibnamefont {Durst}},\ }\href@noop {} {\bibfield
  {journal} {\bibinfo  {journal} {J. Colloid Int. Sci.}\ }\textbf {\bibinfo {volume} {172}},\
  \bibinfo {pages} {147} (\bibinfo {year} {1995})}\BibitemShut {NoStop}%
\bibitem [{\citenamefont {Dalbe}\ \emph {et~al.}(2011)\citenamefont {Dalbe},
  \citenamefont {Cosic}, \citenamefont {Berhanu},\ and\ \citenamefont
  {Kudrolli}}]{dalbe2011aggregation}%
  \BibitemOpen
  \bibfield  {author} {\bibinfo {author} {\bibfnamefont {M.-J.}\ \bibnamefont
  {Dalbe}}, \bibinfo {author} {\bibfnamefont {D.}~\bibnamefont {Cosic}},
  \bibinfo {author} {\bibfnamefont {M.}~\bibnamefont {Berhanu}}, \ and\
  \bibinfo {author} {\bibfnamefont {A.}~\bibnamefont {Kudrolli}},\ }\href@noop
  {} {\bibfield  {journal} {\bibinfo  {journal} {Phys. Rev. E}\ }\textbf
  {\bibinfo {volume} {83}},\ \bibinfo {pages} {051403} (\bibinfo {year}
  {2011})}\BibitemShut {NoStop}%
\bibitem [{\citenamefont {Peng}\ \emph {et~al.}(2008)\citenamefont {Peng},
  \citenamefont {Chen}, \citenamefont {Fischer}, \citenamefont {Weitz},\ and\
  \citenamefont {Tong}}]{tong2009}%
  \BibitemOpen
  \bibfield  {author} {\bibinfo {author} {\bibfnamefont {Y.}~\bibnamefont
  {Peng}}, \bibinfo {author} {\bibfnamefont {W.}~\bibnamefont {Chen}}, \bibinfo
  {author} {\bibfnamefont {T.}~\bibnamefont {Fischer}}, \bibinfo {author}
  {\bibfnamefont {D.}~\bibnamefont {Weitz}}, \ and\ \bibinfo {author}
  {\bibfnamefont {P.}~\bibnamefont {Tong}},\ }\href@noop {} {\bibfield
  {journal} {\bibinfo  {journal} {J. Fluid Mech.}\ }\textbf
  {\bibinfo {volume} {618}},\ \bibinfo {pages} {243} (\bibinfo {year}
  {2008})}\BibitemShut {NoStop}%
\bibitem [{\citenamefont {Du}\ \emph {et~al.}(2012)\citenamefont {Du},
  \citenamefont {Liddle},\ and\ \citenamefont {Berglund}}]{Berglund2012}%
  \BibitemOpen
  \bibfield  {author} {\bibinfo {author} {\bibfnamefont {K.}~\bibnamefont
  {Du}}, \bibinfo {author} {\bibfnamefont {J.}~\bibnamefont {Liddle}}, \ and\
  \bibinfo {author} {\bibfnamefont {A.}~\bibnamefont {Berglund}},\ }\href@noop
  {} {\bibfield  {journal} {\bibinfo  {journal} {Langmuir}\ }\textbf {\bibinfo
  {volume} {28}},\ \bibinfo {pages} {9181} (\bibinfo {year}
  {2012})}\BibitemShut {NoStop}%
\bibitem [{\citenamefont {Sickert}\ \emph {et~al.}(2007)\citenamefont
  {Sickert}, \citenamefont {Rondelez},\ and\ \citenamefont
  {Stone}}]{Stone2007}%
  \BibitemOpen
  \bibfield  {author} {\bibinfo {author} {\bibfnamefont {M.}~\bibnamefont
  {Sickert}}, \bibinfo {author} {\bibfnamefont {F.}~\bibnamefont {Rondelez}}, \
  and\ \bibinfo {author} {\bibfnamefont {H.}~\bibnamefont {Stone}},\
  }\href@noop {} {\bibfield  {journal} {\bibinfo  {journal} { Europhys. 	Lett. }\ }\textbf
  {\bibinfo {volume} {79}},\ \bibinfo {pages} {66005} (\bibinfo {year}
  {2007})}\BibitemShut {NoStop}%
\bibitem [{\citenamefont {Chen}\ and\ \citenamefont {Tong}(2008)}]{tong2008}%
  \BibitemOpen
  \bibfield  {author} {\bibinfo {author} {\bibfnamefont {W.}~\bibnamefont
  {Chen}}\ and\ \bibinfo {author} {\bibfnamefont {P.}~\bibnamefont {Tong}},\
  }\href@noop {} {\bibfield  {journal} {\bibinfo  {journal} { Europhys. 	Lett. }\ }\textbf
  {\bibinfo {volume} {84}},\ \bibinfo {pages} {28003} (\bibinfo {year}
  {2008})}\BibitemShut {NoStop}%
\bibitem [{\citenamefont {Wang}\ \emph {et~al.}(2011)\citenamefont {Wang},
  \citenamefont {Yordanov}, \citenamefont {Paroor}, \citenamefont
  {Mukhopadhyay}, \citenamefont {Li}, \citenamefont {Butt}, \citenamefont
  {Wang}, \citenamefont {Yordanov}, \citenamefont {Paroor}, \citenamefont
  {Mukhopadhay}, \citenamefont {Li}, \citenamefont {Bitt},\ and\ \citenamefont
  {Koynov}}]{koynov2011}%
  \BibitemOpen
  \bibfield  {author} {\bibinfo {author} {\bibfnamefont {D.}~\bibnamefont
  {Wang}}, \bibinfo {author} {\bibfnamefont {S.}~\bibnamefont {Yordanov}},
  \bibinfo {author} {\bibfnamefont {H.~M.}\ \bibnamefont {Paroor}}, \bibinfo
  {author} {\bibfnamefont {A.}~\bibnamefont {Mukhopadhyay}}, \bibinfo {author}
  {\bibfnamefont {C.~Y.}\ \bibnamefont {Li}}, \bibinfo {author} {\bibfnamefont
  {H.-J.}\ \bibnamefont {Butt}}, \bibinfo {author} {\bibfnamefont
  {D.}~\bibnamefont {Wang}}, \bibinfo {author} {\bibfnamefont {S.}~\bibnamefont
  {Yordanov}}, \bibinfo {author} {\bibfnamefont {H.}~\bibnamefont {Paroor}},
  \bibinfo {author} {\bibfnamefont {A.}~\bibnamefont {Mukhopadhay}}, \bibinfo
  {author} {\bibfnamefont {C.}~\bibnamefont {Li}}, \bibinfo {author}
  {\bibfnamefont {H.}~\bibnamefont {Bitt}}, \ and\ \bibinfo {author}
  {\bibfnamefont {K.}~\bibnamefont {Koynov}},\ }\href@noop {} {\bibfield
  {journal} {\bibinfo  {journal} {Small}\ }\textbf {\bibinfo {volume} {7}}
  (\bibinfo {year} {2011})}\BibitemShut {NoStop}%
\bibitem [{\citenamefont {Gehring}\ and\ \citenamefont
  {Fischer}(2011)}]{fisher2014}%
  \BibitemOpen
  \bibfield  {author} {\bibinfo {author} {\bibfnamefont {T.}~\bibnamefont
  {Gehring}}\ and\ \bibinfo {author} {\bibfnamefont {T.~M.}\ \bibnamefont
  {Fischer}},\ }\href@noop {} {\bibfield  {journal} {\bibinfo  {journal} {J.
  Phys. Chem. C}\ }\textbf {\bibinfo {volume} {115}},\ \bibinfo {pages} {23677}
  (\bibinfo {year} {2011})}\BibitemShut {NoStop}%
\bibitem [{\citenamefont {Boniello}\ \emph {et~al.}(2015)\citenamefont
  {Boniello}, \citenamefont {Blanc}, \citenamefont {Fedorenko}, \citenamefont
  {Medfai}, \citenamefont {Mbarek}, \citenamefont {In}, \citenamefont {Gross},
  \citenamefont {Stocco},\ and\ \citenamefont {Nobili}}]{boniello2015}%
  \BibitemOpen
  \bibfield  {author} {\bibinfo {author} {\bibfnamefont {G.}~\bibnamefont
  {Boniello}}, \bibinfo {author} {\bibfnamefont {C.}~\bibnamefont {Blanc}},
  \bibinfo {author} {\bibfnamefont {D.}~\bibnamefont {Fedorenko}}, \bibinfo
  {author} {\bibfnamefont {M.}~\bibnamefont {Medfai}}, \bibinfo {author}
  {\bibfnamefont {N.}~\bibnamefont {Mbarek}}, \bibinfo {author} {\bibfnamefont
  {M.}~\bibnamefont {In}}, \bibinfo {author} {\bibfnamefont {M.}~\bibnamefont
  {Gross}}, \bibinfo {author} {\bibfnamefont {A.}~\bibnamefont {Stocco}}, \
  and\ \bibinfo {author} {\bibfnamefont {M.}~\bibnamefont {Nobili}},\
  }\href@noop {} {\bibfield  {journal} {\bibinfo  {journal} {Nature Materials}\
  }\textbf {\bibinfo {volume} {14}},\ \bibinfo {pages} {908} (\bibinfo {year}
  {2015})}\BibitemShut {NoStop}%
\bibitem [{\citenamefont {Kaz}\ \emph {et~al.}(2012)\citenamefont {Kaz},
  \citenamefont {McGorty}, \citenamefont {Mani}, \citenamefont {Brenner},\ and\
  \citenamefont {Manoharan}}]{kaz2012physical}%
  \BibitemOpen
  \bibfield  {author} {\bibinfo {author} {\bibfnamefont {D.~M.}\ \bibnamefont
  {Kaz}}, \bibinfo {author} {\bibfnamefont {R.}~\bibnamefont {McGorty}},
  \bibinfo {author} {\bibfnamefont {M.}~\bibnamefont {Mani}}, \bibinfo {author}
  {\bibfnamefont {M.~P.}\ \bibnamefont {Brenner}}, \ and\ \bibinfo {author}
  {\bibfnamefont {V.~N.}\ \bibnamefont {Manoharan}},\ }\href@noop {} {\bibfield
   {journal} {\bibinfo  {journal} {Nature Materials}\ }\textbf {\bibinfo
  {volume} {11}},\ \bibinfo {pages} {138} (\bibinfo {year} {2012})}\BibitemShut
  {NoStop}%
\bibitem [{\citenamefont {Frenkel}\ and\ \citenamefont
  {Smit}(2002)}]{frenkelsmit}%
  \BibitemOpen
  \bibfield  {author} {\bibinfo {author} {\bibfnamefont {D.}~\bibnamefont
  {Frenkel}}\ and\ \bibinfo {author} {\bibfnamefont {B.}~\bibnamefont {Smit}},\
  }\href@noop {} {\emph {\bibinfo {title} {Undertstanding Molecular
  Simulation}}},\ \bibinfo {edition} {2nd}\ ed.\ (\bibinfo  {publisher}
  {Academic Press},\ \bibinfo {year} {2002})\BibitemShut {NoStop}%
\bibitem [{\citenamefont {SupplementaryInformation}()}]{suppl}%
  \BibitemOpen
  \bibfield  {author} {\bibinfo {author} {\bibnamefont
  {SupplementaryInformation}}\ }\href@noop {} {\ }\BibitemShut {NoStop}%
\bibitem [{\citenamefont {Razavi}\ \emph {et~al.}(2013)\citenamefont {Razavi},
  \citenamefont {Koplik},\ and\ \citenamefont {Kretzschmar}}]{sepideh2013}%
  \BibitemOpen
  \bibfield  {author} {\bibinfo {author} {\bibfnamefont {S.}~\bibnamefont
  {Razavi}}, \bibinfo {author} {\bibfnamefont {J.}~\bibnamefont {Koplik}}, \
  and\ \bibinfo {author} {\bibfnamefont {I.}~\bibnamefont {Kretzschmar}},\
  }\href {\doibase 10.1039/C3SM50210D} {\bibfield  {journal} {\bibinfo
  {journal} {Soft Matter}\ }\textbf {\bibinfo {volume} {9}},\ \bibinfo {pages}
  {4585} (\bibinfo {year} {2013})}\BibitemShut {NoStop}%
\bibitem [{\citenamefont {Pasol}\ \emph {et~al.}(2011)\citenamefont {Pasol},
  \citenamefont {Martin}, \citenamefont {Ekiel-Jezewska}, \citenamefont
  {Wajnryb}, \citenamefont {Blawzdziewicz},\ and\ \citenamefont
  {Feuillebois}}]{Pasol2011}%
  \BibitemOpen
  \bibfield  {author} {\bibinfo {author} {\bibfnamefont {L.}~\bibnamefont
  {Pasol}}, \bibinfo {author} {\bibfnamefont {M.}~\bibnamefont {Martin}},
  \bibinfo {author} {\bibfnamefont {M.~L.}\ \bibnamefont {Ekiel-Jezewska}},
  \bibinfo {author} {\bibfnamefont {E.}~\bibnamefont {Wajnryb}}, \bibinfo
  {author} {\bibfnamefont {J.}~\bibnamefont {Blawzdziewicz}}, \ and\ \bibinfo
  {author} {\bibfnamefont {F.}~\bibnamefont {Feuillebois}},\ }\href@noop {}
  {\bibfield  {journal} {\bibinfo  {journal} {Chem. Engr. Sci.}\
  }\textbf {\bibinfo {volume} {66}},\ \bibinfo {pages} {4078} (\bibinfo {year}
  {2011})}\BibitemShut {NoStop}%
\bibitem [{\citenamefont {Kremer}\ and\ \citenamefont{Grest}(1986)}]{kremer1986}%
  \BibitemOpen
  \bibfield  {author} {\bibinfo {author} {\bibfnamefont {K.}~\bibnamefont     {Kremer}}\ and\ \bibinfo {author} {\bibfnamefont {G. S.}~\bibnamefont 
  {Grest}},\
  }\href@noop {} {\bibfield  {journal} {\bibinfo  {journal} { Phys. Rev. E
   }\ }\textbf
  {\bibinfo {volume} {36}},\ \bibinfo {pages} {3628} (\bibinfo {year}
  {1986})}\BibitemShut {NoStop}%
\bibitem [{\citenamefont {Allen}\ and\ \citenamefont
  {Tildesley}(1986)}]{Allentildesley}%
  \BibitemOpen
  \bibfield  {author} {\bibinfo {author} {\bibfnamefont {M. P.}~\bibnamefont
  {Allen}}\ and\ \bibinfo {author} {\bibfnamefont {D. J.}~\bibnamefont
  {Tildesley}},\
  }\href@noop {} {\emph {\bibinfo {title} {Computer Simulation of
  Liquids}}},\  (\bibinfo  {publisher}
  {Oxford University Press},\ \bibinfo {year} {1986})\BibitemShut {NoStop}%

\end{thebibliography}
%

\newpage

\centerline{{\bf Supplementary Information}a}

\vspace{0.1in}

\begin{figure}[h]
  \begin{center}
  \includegraphics[width=0.75\linewidth]{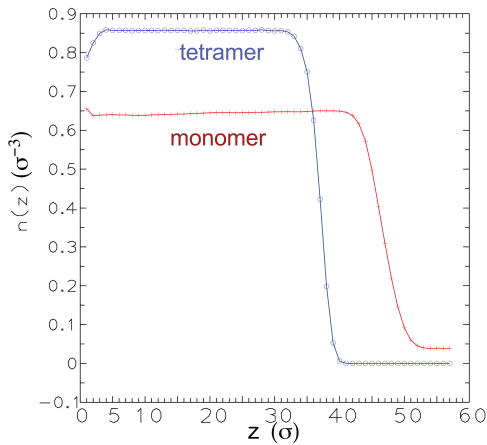}
  \caption{Density profile across the interface for monomeric and 
  tetrameric liquids.}
  \label{densityprofile}
  \end{center}
\end{figure}

\noindent{\bf{Molecular Dynamics Simulation Details}}
\newline
The tetrameric liquid is made of atoms of diameter $\sigma$ and mass $m$ 
which are tethered nto chains by the FENE potential 
\begin{equation}
V_{FENE}(r) = -\frac{1}{2}\, k_F\, r_0^2\, \ln\left(1-{r^2\over r_0^2}\right),
\end{equation}
with parameters $k_F=30\epsilon/\sigma^2$ and $r_0=1.5\sigma$ (following 
Ref.~\cite{kremer1986}).
In addition, all fluid atoms interact by a Lennard Jones interaction with 
energy $\epsilon$ and with an attractive parameter $c=1$. 
The motivation for using tetrameric molecules is to sharpen the interface, 
which would be broader than the particle itself in the monatomic case: 
the interfacial density profiles in the absence of a particle is shown 
in Fig.~\ref{densityprofile}.
The molecules are placed in a simulation cell, which,
as noted in the paper, is in the form of a slab with a free surface
at the top and in contact with a bottom consisting of two layers
of LJ atoms (of identical mass $m$ and diameter $\sigma$), 
attached by linear tether
springs to fcc lattice sites. The solid atoms interact with the
liquid atoms through an LJ interaction with attractive parameter
$c=1$. Periodic boundary conditions are imposed in the lateral
directions. The simulation cell  is a box of side 60.3 $\sigma$.
The temperature is fixed by a Nose-Hover thermostat 
at $0.8 \epsilon/k_{B}$, which provides a liquid/vapor equilibrium
with  bulk density iequal to  $0.857\sigma^{-3}$ while the solid
base is 1.2 times denser to prevent leakage. The equations of motion
are integrated using predictor-corrector methods. The particle is
carved from a rigid cubic lattice of atoms (again mass $m$ and diameter
$\sigma$) to form a colloidal particle  with a rough surface. The colloid
atoms interact with the tetrameric liquid through an LJ interaction
with attractive parameter $c$.  The particle motion is obtained by
computing the net force and torque exerted on it by fluid atoms,
and translating and rotating it according to the Newton and Euler
equations.  Quaternion variables \cite{Allentildesley} are used to
specify the orientation. There are 110,608 atoms in the liquid, 3600 in the
solid base and 1736 in the particle, and the simulation box is a cube of
side 60.3$\sigma$.  If the parameters in the potential are
given values appropriate for argon, the particle radius is about 2.5 nm and
the intrinsic time unit $\tau=2.12$ ps.
Most numerical measurements reported
here involve only the position of the center of the particle or the net
force exerted on its atoms by the fluid. For 
the density field (illustrated in Fig. 1 of article), we employ 
a two-dimensional array of sampling bins taken in the rest frame of the
particle, and record the average occupancy over a 10$\tau$ interval. 
\\
\noindent{\bf{Particle Orientation Details}}

\begin{figure}[t]
  \begin{center}
  \includegraphics[width=0.75\linewidth]{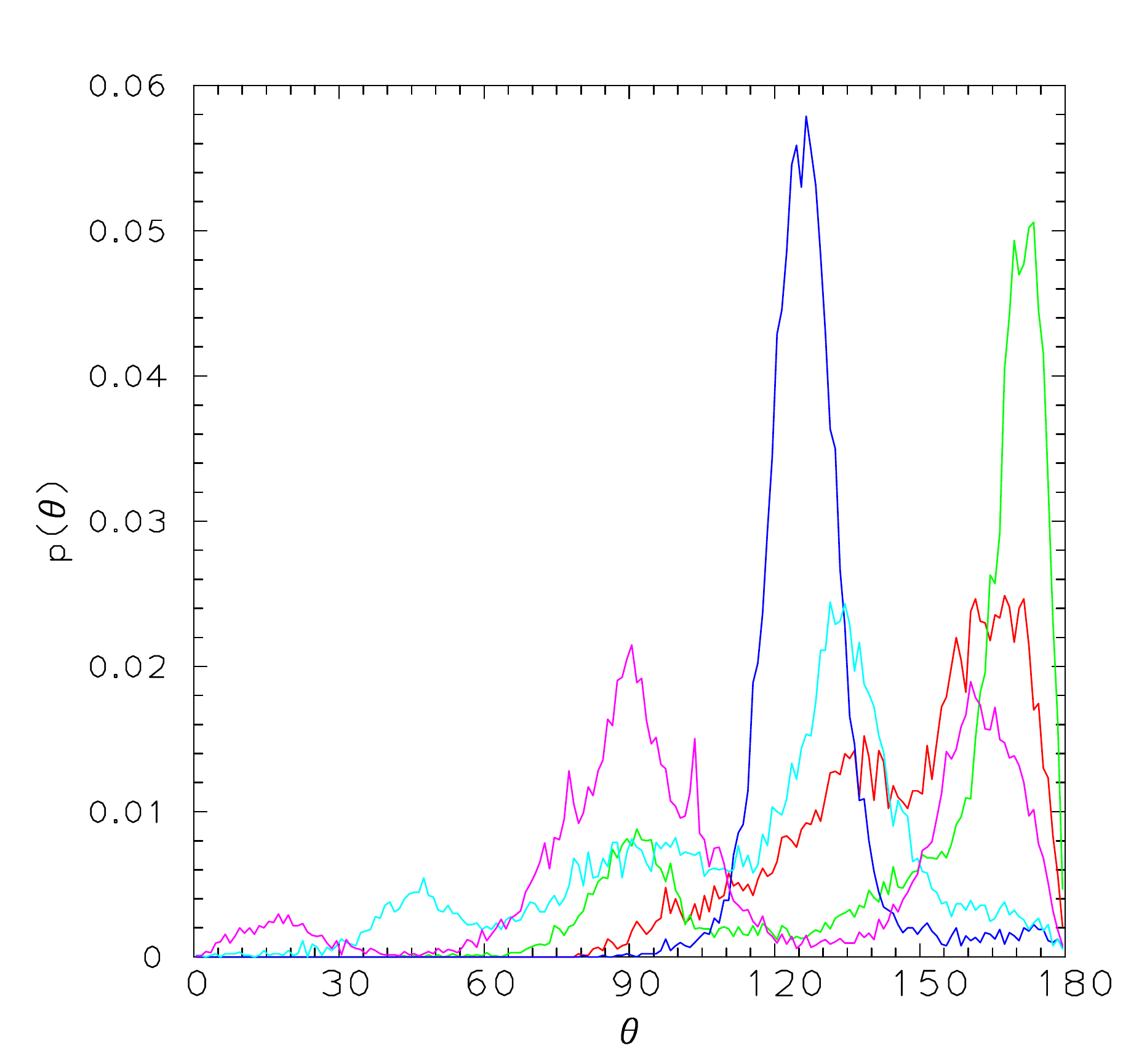}
  \caption{Probability distributions for the orientation angle $\theta$
  normal to the interface, for different wettabilities. The color code is the
  same as in Fig. 2 and 3 of the article.}
  \label{fig:theta}
  \end{center}
\end{figure}
To elaborate on the orientational behavior of the particle during the 
diffusion runs, we have averaged over the different realizations to obtain 
the probability
distribution functions for the two angles. Unfortunately the results must
be regarded as transient since the simulations do not run long enough to
fully sample the angular distributions. The characteristic
rotational diffusion time $T_r$ is the inverse of the rotational 
diffusivity, and using the Stokes-Einstein relation again, 
$T_r = 1/D_r = \xi_r/k_BT =8\pi\mu R^3/k_BT = 8.33\times 10^4\tau$,
which rather exceeds the duration of the simulations here.  
The transient probability distribution functions for $\theta$ over a
1000$\tau$ interval for different wettabilities are shown in 
Fig.~\ref{fig:theta}. We see either none or one or two peaks, 
at different angles in different wettabilities, with no obvious pattern. 
The corresponding
distribution for the in-plane angle $\phi$ are similarly devoid of a clear
pattern and generally are much noisier versions of this figure. 

\end{document}